\documentstyle[newarcrc,fleqn,epsfig]{article}

\title{MERLIN observations of GRS 1915+105 : \\
a progress report}

\author{R.~P.~Fender\address{University of Amsterdam, The Netherlands},
        S.~T.~Garrington\address{University of Manchester, Jodrell
        Bank, UK}, 
	D.~J.~McKay$^{\rm b,}$\address{Joint Institute for VLBI in
Europe, The Netherlands}, 
	T.~W.~B.~Muxlow$^{\rm b}$, \\
	G.~G.~Pooley\address{Mullard Radio Astronomy Laboratory, UK}, 
	R.~E.~Spencer$^{\rm b}$,
        A.~M.~Stirling$^{\rm b}$, 
	E.~B.~Waltman\address{Remote Sensing Division, NRL, USA}}

\begin{document}
\maketitle

\begin{abstract}

We present a progress report on MERLIN radio imaging of a radio
outburst from GRS 1915+105. The major ejection occurred at the end of
a $\sim 20$ day `plateau' state, characterised by low/hard X-ray
fluxes and a relatively strong flat-spectrum radio component.
Apparent superluminal motions have been mapped with unprecedented
resolution, and imply higher velocities in the jet than previously
derived.

\end{abstract}

\section{Introduction and Observations}

We have observed the superluminal X-ray transient (Mirabel \&
Rodriguez 1994, hereafter MR94) with the MERLIN array in 1997 October
-- November following a major radio outburst.

The MERLIN observations were made with an angular resolution of 40
mas, five times better than previous VLA mapping, at a frequency of 5
GHz, on ten occasions over a period of 12 days. The radio flux from
GRS 1915+105 was also simultaneously monitored at 2 \& 8 GHz with the
Green Bank Interferometer (GBI) and at 15 GHz with the Ryle Telescope
(RT).  Over the same period the system was monitored in soft X-rays
with the Rossi XTE satellite.

\section{X-ray and radio state}

Prior to our MERLIN observations GRS 1915+105 had clearly been in 
an unusual state in both X-rays and radio, summarised in Fig. 1.
Approximately 25 days prior to our first epoch of mapping, the source
had undergone a significant radio flare, and then entered a `plateau'
state. This state is characterised by relatively bright and stable
radio emission with a flat / inverted spectrum, and low but
persistent X-ray emission with a hard spectrum. Previous 
monitoring reported at 15 GHz with the RT in Pooley \& Fender (1997)
and at lower frequencies in Bandyopadhyay et al. (1998) confirm the
association between the plateau state and relatively bright radio, and
perhaps infrared, emission.

\begin{figure}
\centerline{\epsfig{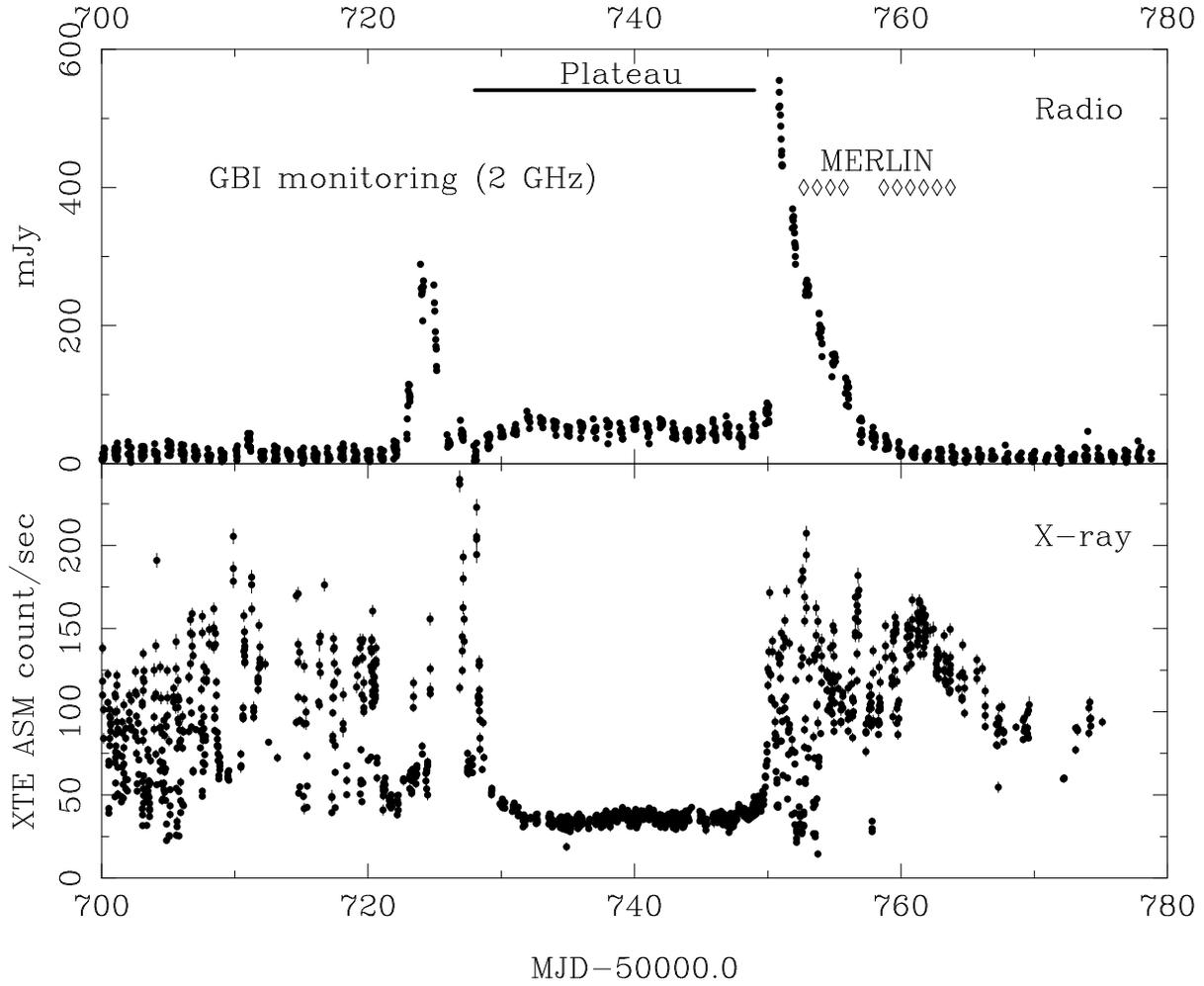}}
\caption{Radio (GBI) and X-ray (XTE) monitoring around the time of the
MERLIN observations of GRS 1915+105. Around MJD 50725 a significant
radio flare signals the beginning of the plateau state, characterised
by relatively bright and stable flat-spectrum radio emission and
steady, hard-spectrum emission in X-rays. The flare at the end of the plateau
period triggered our MERLIN observations, indicated by diamonds on the
top panel. Images from the first four epochs of MERLIN observations
are presented in Fig. 2.}
\end{figure}

\section{Superluminal ejections}

\begin{figure}
\centerline{\epsfig{file=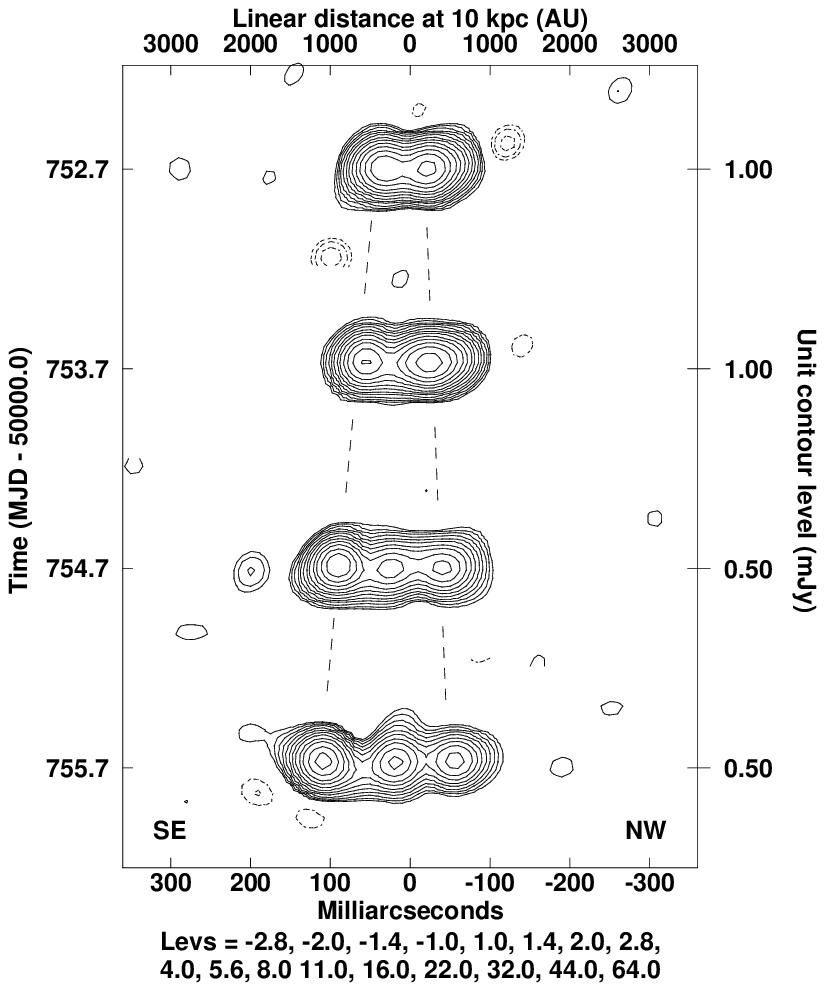,width=14cm,clip}}
\caption{Radio maps of GRS 1915+105 from the first
four epochs of our MERLIN observations. The observing frequency is 5
GHz (6 cm) and the angular resolution 40 mas. In the first two epochs
the core and receding component are blended together; by the third
epoch core and approaching (SE) and receding (NW) components are
clearly resolved. The proper motions are significantly higher than
those reported in MR94 (see text). The unit contour level for each
epoch is indicated on the right side of the figure, and the maps have
been rotated clockwise by $\sim 145$ degrees for clarity.}
\end{figure}

As indicated in Fig. 1, a sequence of ten MERLIN high-resolution radio
maps were made of GRS 1915+105. The first four of these are presented
in Fig. 2, revealing the clear expansion of the source on daily
timescales.  The jet sidedness is consistent with that reported in
MR94, and we are able to track the proper motions of three approaching
and one receding components over the entire set of images. The
position angle of the ejections, approximately 145 degrees, is also very
similar to that reported in MR94.

All ejections are consistent with ballistic motions to better than
10\%, and we find best-fits to the proper motions of :

\[
\mu_{\rm app} = 23.6 \pm 0.5
\]

and

\[
\mu_{\rm rec} = 10.0 \pm 0.5
\]

mas. d$^{-1}$, significantly greater than those reported in MR94.  All
fits are good, with $\chi^2_{\rm red} \leq 1$. The proper motion of
$17.6 \pm 0.4$ mas. d$^{-1}$ reported by MR94 for the approaching
component can be ruled out for this ejection; fixing the proper motion
to this value does not give an acceptable fit to the data.

Initial analysis of the proper motions, under the standard assumption
of an intrinsically symmetric ejection, suggests that we are measuring
significantly greater velocities (by $\geq 10$\%) than MR94, at a
similar angle to the line of sight (most likely in the range 60 -- 70
degrees). 

Again assuming a symmetric ejection, we can use the proper motions to place
limits on the distance to GRS 1915+105. At the most extreme, given the
measurement uncertainties, the distance must be $\leq 13.6$ kpc. A
more likely upper limit is $\leq 11.2$ kpc, i.e. significantly closer
(although within their estimated errors) than the best estimate of
12.5 kpc used by MR94.

\section{Conclusions}

We have presented preliminary results from a major set of MERLIN
observations of GRS 1915+105 simultaneous with multiwavelength radio
and X-ray monitoring. While the nature of the plateau state itself
remains unclear, it is now certain that the radio flaring associated
with the end of the state {\bf is} associated with relativistic
ejections. 

These ejections, in particular the approaching component, have
significantly higher proper motions than those recorded by MR94, most
simply interpreted as higher intrinsic velocities at a similar angle
to the line of sight. We also find that GRS 1915+105 is likely to be
significantly closer than the best estimate of 12.5 kpc of MR94.

More details of the observations, including spatially resolved linear
polarisation maps, and their interpretation, will be found in a paper
to be submitted to MNRAS.

\section*{Acknowledgements}

We acknowledge great efforts by many people, mostly at Jodrell Bank,
which helped to make these observations a success. In particular we
would like to thank Peter Thomasson, Shane McKie and Richard Ogley.
MERLIN is operated by the University of Manchester at the Nuffield
Radio Astronomy Laboratories, Jodrell Bank, on behalf of the Particle 
Physics and Astronomy Research Council. We acknowledge also the use of
public data from XTE ASM and GBI. RPF acknowledges support from EC
Marie Curie Fellowship ERBFMBICT 972436, and DJM from EC TMR-LSF
contract No. ERBFMGECT 950012.

\end{document}